\documentclass{ptapap}

\author{Karolina Bąkowska}[CAMK,OSU]
\author{Arkadiusz Olech}[CAMK]
\author{Remigiusz Pospieszyński}[COM]
\affil[CAMK]{Nicolaus Copernicus Astronomical Center\\
  Bartycka 18, 00--716 Warszawa, Poland}
\affil[OSU]{Fulbright Visiting Scholar, The Ohio State University\\  Dept. of Astronomy, 140 W. 18th Ave, Columbus, OH 43210, USA}
\affil[COM]{Comets and Meteor Workshop\\ Bartycka 18, 00-716 Warszawa, Poland}

\title{MN Dra - a SU UMa-type star during its September 2013 superoutburst}

\begin{document}

\maketitle

\begin{abstract}

We report CCD photometry of the cataclysmic variable star MN Draconis. During the season of August-September 2013, one normal outburst and one superoutburst were detected. In the light curves of MN Dra clear superhumps were present during September 2013 superoutburst. That fact confirms that the star is a member of SU UMa class.

\end{abstract}

\section{Introduction}

Cataclysmic variable stars are close pairs containing a white dwarf (the primary) and usually a the Roche-lobe filling main-sequence star (the secondary). The period gap (from 2 to 3 hours) is an orbital period range where there is a significant dearth of active cataclysmic variables. 

SU UMa-type dwarf novae are one of the subclasses of cataclysmic stars, characterized by the short orbital periods (below $2.5$ hours). In the light curves of the SU UMa novae one can observe a sudden rise of brightness called normal outbursts or superoutburts. Outburts have shorter duration and lower brightness than superoutbursts during which tooth-shaped oscillations called superhumps are observed \citep{2001cvs..book.....H}.

MN Dra is an SU UMa-type dwarf nova in the period gap. \cite{2010OAP....23...98S} detected two superoutbursts and five outbursts of MN Dra during 77 nights of observations in August-November 2009.  \cite{2014PASJ...66...90K} observed MN Dra during the 2012 July-August and the 2013 November superoutbursts. Our team performed an analysis of the superhump period during September 2013 superoutburst.

\section{Observations}

We performed the observation campaign of MN Dra during 10 nights from 2013 August 12 to September 10. All data were gathered at the Ostrowik station of Warsaw Astronomical Observatory. 

\begin{figure}
\centering
   \includegraphics[width=0.8\textwidth]{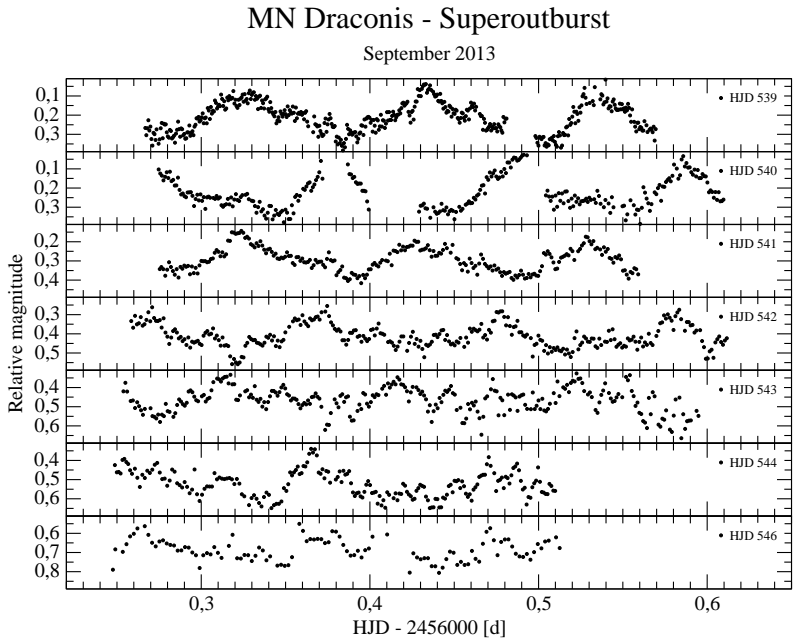}
   \caption{The MN Dra superhumps during its September 2013 superoutburst.}
   \label{fig:Superhumps}
\end{figure}

MN Dra was monitored in a clear filter ("white" light).Bias, dark current and flat-field corrections were made using the IRAF\footnote{IRAF is distributed by the National Optical Astronomy Observatory, which is operated by the Association of Universities for
Research in Astronomy, Inc., under a cooperative agreement
with the National Science Foundation.} package and profile photometry with DAOPHOTII package \citep{1987PASP...99..191S}. Fig.\ref{fig:Superhumps} shows the light curves of MN Dra during 7 nights of the September 2013 superoutburst. The superhumps are clearly visible in each run for all displayed nights.

\section{Results}

The $O-C$ diagram is an excellent tool to check the stability of the superhump  period and to determine its value. We analysed timings of primary maxima. In total, we were able to determine 16 moments of maxima. The least squares linear fit to the maxima timings gives the following ephemeris:

\begin{equation}
{\rm HJD_{\rm max}} = 2456539.3276(6) + 0.10496(2) \times E		
\label{Eq1}
\end{equation} 

\begin{figure}
\centering
   \includegraphics[width=0.8\textwidth]{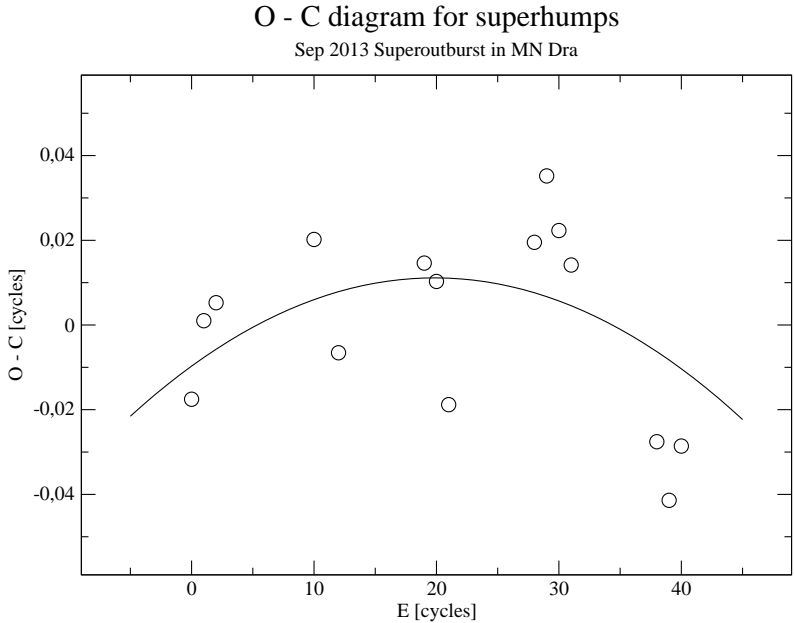}
   \caption{ The $O-C$ diagram for superhumps maxima of MN Dra detected during its September 2013 superoutburst.}
   \label{fig:OC}
\end{figure}

The $O-C$ values corresponding to the ephemeris \ref{Eq1} are shown in Fig.\ref{fig:OC}. Additionally, the second-order polynomial fit was calculated for the moments of maxima. The following ephemeris was obtained:

\begin{equation}
{\rm HJD_{\rm max}} = 2456539.3266(7) + 0.10518(8) \times E - 5.6(1.9) \times 10^{-6}	\times E^2		
\end{equation}

Due to the poor agreement between second-order polynomial fit and the $O-C$ values of the moments of maxima, we cannot confirm the decreasing trend of the superhump period postulated by \cite{2010OAP....23...98S}. 

\section{Summary}

Our team observed MN Dra during its August 2013 outburst and September 2013 superoutburst. Based on the data presented by \cite{2010OAP....23...98S} and \cite{2014PASJ...66...90K}, we calculated the supercycle length for $~61$ days. Due to the fact that MN Dra is another example, e.g. \cite{2014AcA....64..337B}, of an active SU UMa dwarf nova from the period gap which poses a serious problem for the theory of superhumps and superoutbursts, we plan further analysis of this cataclysmic variable star.

\acknowledgements{We acknowledge generous allocation of the Warsaw Observatory 0.6-m telescope time. Project was supported by  Polish National Science Center grants awarded by decisions DEC-2012/07/N/ST9/04172 and DEC-2015/16/T/ST9/00174 for KB.}

\bibliographystyle{ptapap}
\bibliography{ptapapdoc}

\end{document}